# Transition from steady to oscillating convection rolls in Rayleigh-Bénard convection under the influence of a horizontal magnetic field

**J.C. Yang[1,2], T. Vogt[1], and S. Eckert[1]**

[1] MHD Division, Helmholtz-Zentrum Dresden-Rossendorf (HZDR), 01328 Dresden, Germany
[2] State Key Laboratory for Strength and Vibration of Mechanical Structures, School of Aerospace, Xi'an Jiaotong University, Xi'an, Shaanxi 710049, China

## Abstract

In this study we consider the effect of a horizontal magnetic field on the Rayleigh-Bénard convection in a finite liquid metal layer contained in a cuboid vessel ($200{\times}200{\times}40$ mm$^3$) of aspect ratio $\varGamma = 5$. Laboratory experiments are performed for measuring temperature and flow field in the low melting point alloy GaInSn at Prandtl number $Pr \approx 0.03$ and in a Rayleigh number range $2.3{\times}10^4 \leq Ra \leq 2.6{\times}10^5$. The field direction is aligned parallel to one pair of the two side walls. The field strength is varied up to a maximum value of 320 mT ($Ha = 2470$, $Q = 6.11{\times}10^6$, definitions of all non-dimensional numbers are given in the text). The magnetic field forces the flow to form two-dimensional rolls whose axes are parallel to the direction of the field lines. The experiments confirm the predictions made by Busse and Clever (J. Mécanique Théorique et Appliquée, 1983 [1]) who showed that the application of the horizontal magnetic field extends the range in which steady two-dimensional roll structures exist ('Busse balloon') towards higher $Ra$ numbers. A transition from the steady to a time-dependent oscillatory flow occurs when $Ra$ exceeds a critical value for a given Chandrasekhar number $Q$, which is also equivalent to a reduction of the ratio $Q/Ra$. Our measurements reveal that the first developing oscillations are clearly of two-dimensional nature, in particular a mutual increase and decrease in the size of adjacent convection rolls is observed without the formation of any detectable gradients in the velocity field along the magnetic field direction. At a ratio of $Q/Ra \approx 1$, the first 3D structures appear, which initially manifest themselves in a slight inclination of the rolls with respect to the magnetic field direction. Immediately in the course of this, there arise also disturbances in the spaces between adjacent convection rolls, which are advected along the rolls due to the secondary flow driven by Ekman pumping. The transition to fully-developed three-dimensional structures and then to a turbulent regime takes place with further lowering $Q/Ra$.

## I. INTRODUCTION

Natural convection is responsible for large-scale fluid motion in nature and industrial applications. Such convective systems are studied on a laboratory scale in generic configurations, of which Rayleigh Bénard convection (RBC), which considers thermally driven convection between two horizontal plates kept at constant temperatures, is the most fundamental and classical configuration (see for example: Bodenschatz et al. [2], Ahlers et al. [3], F. Chillà and J. Schumacher [4]). The convective flow and the related heat and momentum transport properties are governed by two dimensionless numbers, the Rayleigh number, $Ra = \alpha g \varDelta T H^3 / \kappa \nu$ and the Prandtl number, $Pr = \nu / \kappa$, where $g$, $\varDelta T$, and $H$ denote the acceleration by gravity, vertical temperature differences, and height of the fluid layer. The thermo-physical properties of the liquid are described by the thermal expansion rate $\alpha$, the thermal diffusivity $\kappa$, and kinematic viscosity $\nu$, respectively. When considering confined flows, the shape and dimensions of the fluid container also become relevant. Here, the aspect ratio $\Gamma = L/H$ is another dimensionless variable that comes into play, while $L$ stands for the characteristic horizontal dimension of the fluid vessel. Convective flows of astrophysical relevance concern fluids with a small $Pr$ number and develop under the influence of magnetic fields [5]. The impact of the magnetic field is represented by the Chandrasekhar number Q = $B^2L_B^2\sigma/\rho\nu$ (Hartmann number Ha squared) which is a measure for the ratio of the electromagnetic force to the viscous force, where $\sigma$ and $\rho$ are the electrical conductivity and the density of the fluid, respectively. $B$ denotes the strength of the applied magnetic field and $L_B$ the length of the fluid layer in magnetic field direction.

According to the pioneering stability theory provided by [6,7], the onset of convection in infinite liquid layers is independent of $Pr$, whereas the further development of convective motion with increasing $Ra$ is also notably dependent on $Pr$. When a critical value of the temperature gradient $\Delta T$ is exceeded, the originally motionless fluid layer becomes unstable in favor of the formation of a periodic pattern of steady two-dimensional (2D) convection rolls with the characteristic wave number $k$. Stability analysis also predicts the existence of a domain in the $\Delta T - k$ plane, well-known as the so-called 'Busse balloon', within which the emerged 2D roll structure remains stable. A continuous increase of $Ra$ leads to leaving the 'Busse balloon' by the transition from the steady 2D rolls to time-dependent three-dimensional (3D) convection. The emerging time-dependent motion is known as oscillatory instability [8] and becomes manifest as traveling waves in the direction of the roll axis [9-10]. Krishnamurti and Howard [11] described the development of the supercritical flow with increasing Ra number as a sequence of several stages, starting with steady 2D rolls via steady and time-dependent 3D flows to turbulence. It appears that the transition process from steady convection to thermal turbulence is more rapid and immediate the smaller the $Pr$ number becomes, that means that the dimensions of the balloon are considerably reduced as $Pr$ decreases [8]. In liquid metals at low $Pr$ conditions, $Pr = O(10^{-2})$, the cross-section of the balloon actually becomes so small that quantitative investigations of stable convection rolls in this range are incredibly difficult.

A realistic option to expand the 'Busse balloon' is to apply a stabilizing force in the form of a stationary magnetic field. Busse and Clever [1] found that the region of stable rolls is considerably increased by a longitudinal magnetic field along the roll axes. Such a delay in the onset of oscillatory convection was also observed in experiments conducted in a mercury layer by Fauve et al. [10]. The linear stability theory shows that applying a horizontal magnetic field to infinite fluid layers does not affect the critical value $Ra_c$ for the onset of convection [12]. Steady solutions for a magnetic field aligned parallel to the roll axes are identical to those of the non-magnetic case [1], however, the magnetic field suppresses three-dimensional disturbances. The mechanism of the so-called magnetic damping consists in the conversion of mechanical into thermal energy by means of Joule dissipation. However, the magnetic field effect is anisotropic causing a propagation of vorticity and linear momentum along the field lines [13]. The effect of a magnetic field on originally 3D flows becomes apparent by a complete reshaping of the flow into a 2D structure. MHD turbulence is a prominent example, where the magnetic field effectively inhibits the development of 3D fluctuations, thus de facto increasing the intensity of 2D structures aligned along the magnetic field lines [14,15]. In the particular case of liquid metal RBC, a horizontal magnetic field effectively prevents 3D flow pattern from absorbing the energy introduced by the buoyancy. Instead, the motion of the 2D convection rolls is intensified. This is verified by an increase of the both the roll circulation velocity and the frequency of the oscillations [10]. Consequently, this is also associated with growing effective thermal conductivity, namely with an improvement of the heat transfer [16, 17].

Unbounded flows, that have developed ideal 2D structures under the impact of an external DC magnetic field, are then no longer affected by electromagnetic forces. In finite fluid layers delimited by side walls, boundary-layer effects become relevant and determine the magnetic damping. In case of electrical non-conducting walls, the loops of induced electric currents close exclusively within thin Hartmann boundary layers forming along the walls, which confine the fluid in direction of the magnetic field. Joule and viscous dissipation now primarily occur within these Hartmann layers causing the so-called Hartmann braking [18]. Burr and Müller [17] accounted for this effect by establishing a neutral stability curve for the onset of convection depending on $Q$ and predicted a growing wave number of the 2D convection rolls with increasing $Q$. Furthermore, Burr and Müller [17] conducted laboratory experiments using Sodium-Potassium eutectic alloy, $Na_{22}K_{78}$ (Pr $\approx 0.02$). The existence of coherent time-dependent quasi-two-dimensional flow at high field intensities and moderate $Ra$ was confirmed by measuring the temperature at three positions along the magnetic field. The temporal dynamics of the convective flow became dominated by a few superimposed or even one single frequency. Moreover, they were able to identify a parameter range, where stabilizing of increasingly non-isotropic convection pattern goes along with an increase of the heat transfer. However, due to the non-availability of techniques for direct flow measurement in this experiment, it was not possible to resolve the structure of the flow regimes, analyze their time behavior and study the transition from stationary to time-dependent convection in detail.

Besides the question to what extent the magnetic field will be able to expand the stability range of steady-state convection, an interesting aspect is also to investigate what kind of time-dependent flow is established when leaving the 'Busse balloon' with increasing $Ra$. Numerical simulations for buoyancy-driven convection in an infinite fluid layer predict a transition from steady convection to traveling waves via Hopf bifurcation [19]. In particular, it involves a transverse time-dependent fluctuation of the rolls. Although there is widespread agreement on the fact that a longitudinal magnetic field suppresses these 3D effects and delays their onset, it is not yet fully understood whether quasi 2D oscillations, which are more compatible with the magnetic field, may also occur. Oscillating flow regimes have been observed by Yanagisawa et al. [20,21] in a liquid gallium layer of aspect ratio $\Gamma = 5$ under the impact of a weak magnetic field. The authors suggested a regime diagram in which these oscillatory flows separate the area of stationary flow regimes from the turbulent convection. Apparently, there are several oscillatory regimes differing quite clearly from one another in terms of amplitude and dominant frequencies. However, the respective flow topology was not further investigated in these studies. A more recent investigation provides more detailed knowledge of the flow structures in the same experimental geometry [22]. Linear velocity profiles were measured by the ultrasonic Doppler technique along measuring lines parallel and perpendicular to the magnetic field, respectively. In addition, the parameter range has been significantly increased by applying stronger magnetic fields. The flow measurements provide first indications for the assumption that the oscillations emerging just after the transition from stationary convection rolls to time-dependent flows could have a two-dimensional character as long as the magnetic field is still sufficiently strong ($Q/Ra > 1$). Distinct 3D effects occur with further lowering of the ratio $Q/Ra$. The authors assumed that the oscillations begin as soon as the rolls become deformed, i.e. that their cross-sectional areas start to deviate from the original circular shape. A rotation of these non-axisymmetric structures should be recorded by the sensors as oscillations coupled with the roll's turnover time. Another recent paper [23] considers the effect of the flow structure on the heat and momentum transfer in the same experimental setup as used in [22]. The authors succeeded in proving that the non-monotonic behavior of heat transport with increasing Q number, already observed by Burr and Müller [17], is indeed due to transitions between different flow regimes. The amplification of the flow is even so distinct that the velocity components perpendicular to the magnetic field reach the value of the free-fall velocity [23]. They distinguish between three main flow regimes: two 3D regimes showing a cellular structure or an unstable roll configuration and a quasi-2D configuration of oscillating convection rolls. The present study focuses on the parameter subrange where a stable configuration of convection rolls is found. These stable rolls are aligned with the magnetic field direction and fill the entire cell width, so it can be assumed that the induced currents close mainly within the Hartmann layer [14]. It was observed that this roll structure appears completely laminar and stationary at very high magnetic fields, but starts regular oscillations with a decrease of the magnetic field strength [22]. An open question is whether these oscillating structures can be described as quasi-2D dimensional in a strict sense, and in what form a transition to a three-dimensional flow occurs.

Some general remarks on the definition of the term quasi-two-dimensional flow and its use in this paper should be made at the beginning. The decisive feature of any two-dimensional flow is its strict orientation along a selected direction in space (here parallel to the magnetic field), i.e. the velocity component and all possible velocity gradients along the magnetic field lines approach zero. Flows in infinitely extended layers can reach such an ideal two-dimensional state, but in real experiments quasi-two-dimensional flows are formed under the influence of the side walls. The flow structure is two-dimensional in its core, but large gradients of velocity in the direction of the magnetic field occur in the boundary layers (Hartmann layers). Moreover, any rotating flow over a solid wall creates a secondary flow driven by Ekman pumping [24]. In this sense, we always have a flow along the magnetic field, whose intensity in relation to the rotation speed of the rolls depends on the magnetic field strength [22]. This situation corresponds to the so-called Bödewadt-Hartmann problem, for which Davidson and Potherat [25] provided a detailed theoretical analysis. A prominent feature is the formation of combined Bödewadt-Hartmann layers at the wall perpendicular to the magnetic field. This phenomenon was also investigated for a single electrically-driven vortex between two parallel walls [26,27]. Applying linear stability analysis [6,7,17] showed that after the onset of convection the flow is supposed to be stationary. Oscillating flow structures arise from secondary bifurcation, which is evoked with increasing $Ra$ number. We expect that the first oscillating structures, which occur when $Q$ is still sufficiently high, will have a two-dimensional character. As soon as the magnetic field loses its dominant role with decreasing $Q$, three-dimensional disturbances will develop.

This study is a continuation of the work of Vogt et al. [22]. We have significantly increased the number of measurements for various parameter combinations ($Ra$, $Q$) to adequately investigate the details of the transitions from a stationary roll regime to quasi-2D oscillations and from these 2D flow patterns to 3D flows. A major objective is to elucidate the details of the topology of the respective oscillating structure. For this purpose, the evaluation methods for the interpretation of the flow measurements were further refined compared to our previous study. In contrast to the paper by Vogt et al. [22], however, the definition of the Chandrasekhar number has been modified. As characteristic length we do not consider the height of the convection cell, but the length parallel to the magnetic field, $L_B$. The definition of $Q$ using the height $H$ goes back to the studies of Burr and Müller, who used the same definition of the Hartmann number $Ha$ with $H$ as characteristic length for their investigations both in the vertical [28] and the horizontal magnetic field [17]. In our opinion, the use of the horizontal dimension of the cell is better suited for the case of a horizontal magnetic field, since the effect of Hartmann braking scales with the dimension of the flow domain in magnetic field direction [29,30].

## II.    EXPERIMENTAL SETUP

The experimental setup is based on a specific configuration first applied at Hokkaido University in Sapporo [21]. The actual version of the convection cell was designed at the Helmholtz Zentrum Dresden-Rossendorf (HZDR) and has already been used in some recent studies [22,31,32]. The liquid container has a square cross-section with side lengths $L$ of 200 mm each and a height $H$ of 40 mm, as shown in Fig. 1(a). It is bounded by copper plates at the top and bottom, in which a network of channels is incorporated and connected to a water circulation system. The operation of thermostatic baths and careful control of the water flow rates are supposed to ensure isothermal boundary conditions with set temperature gradients. The side walls are made of electrical non-conducting polyvinyl chloride (PVC). The convection cell is encased in 30 mm closed cell foam to minimize heat loss.

The fluid vessel is filled with GaInSn eutectic ($Ga_{67}In_{20.5}Sn_{12.5}$); a ternary metal alloy ($Pr = 0.03$) which has a melting point lower than room temperature. Thermophysical properties of GaInSn have been published by Morley et al. [33] and Plevachuk et al. [34]. The rectangular convection cell is situated in an almost uniform magnetic field whose spatial homogeneity at the maximum field strength of 700 mT is better than 5 %. The magnetic field direction is aligned parallel with respect to one of the side wall pairs of the vessel. The magnetic field was generated by an electromagnet, in which two water-cooled copper coils are arranged on a large iron yoke. The distance between the pole shoes can be flexibly adapted to the dimensions of the convection cell.

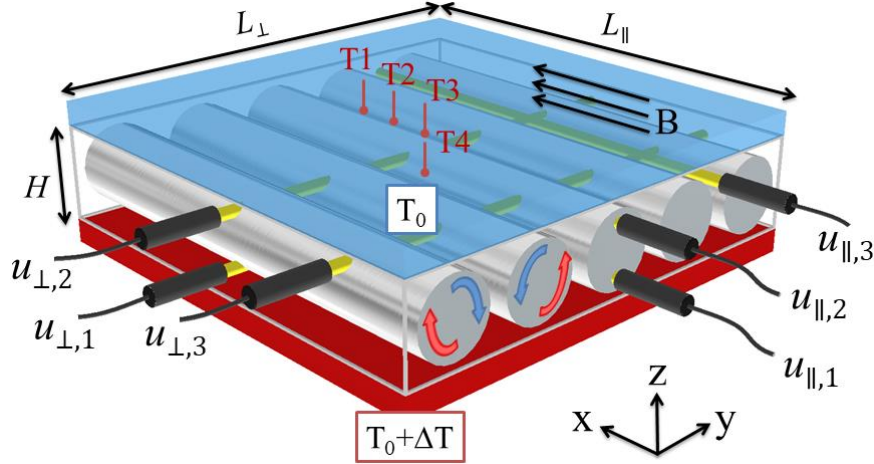

FIG. 1. Schematic drawing of the experimental setup showing the convection cell and sensor configuration

The temperature difference over the height of the convection cell, on which the calculation of $Ra$ is based, is determined using nine thermocouples in both the top and bottom copper plates. Another 4 thermocouples are positioned 3 mm below the upper copper plate to monitor local temperatures in the thermal boundary layer of the fluid. The arrangement of the sensors is illustrated in Fig. 1. Thermocouple T3 is located exactly in the middle of the cross-sectional area. Two sensors T1 and T2 are placed at intervals of 20 mm in the direction of the magnetic field, while another sensor T4 is installed also at 20 mm distance perpendicular to it.

Over the last two decades, the ultrasound Doppler velocimetry (UDV) has been developed into an efficient and powerful instrument for measuring flow velocities in liquid metals [35-37]. This technique has also found its application for investigations on liquid metal convection (for example Mashiko et al. [38], Tsuji et al. [39], Yanagisawa et al. [20] and Vogt et al. [40]). Recently, Zürner et al. [41] performed combined measurements of temperature and velocity to reconstruct the LSC structure in a $\Gamma = 1$ cylinder. The measurement instrument DOP 3010 (Signal Processing S.A.) is employed within this study. Ultrasonic transducers with an emitting frequency of 8 MHz and a piezo element having an active diameter of 5 mm are chosen. The measuring volume of the velocity profiles along the propagation direction of the ultrasound consists of a sequence of disc-shaped segments per measuring point, each with a dimension of approx. 1 mm in axial direction and 5 mm in lateral direction. A time resolution of 1 Hz is realized with a sequential scanning of all sensors. Six transducers are mounted horizontally inside holes drilled into the two side-walls of the vessel. A suitable preparation of the front end of the transducers and their direct contact with the liquid metal ensures a low-loss transmission of the ultrasonic wave into the liquid to be measured. Four transducers ($u_{\perp,2}, u_{\perp,3}, u_{\parallel,2}$ and $u_{\parallel,3}$) in Fig. 1(b) are installed at a distance of 10 mm to the top plate ($z = 3/4$ $H$) and the other two transducers ($u_{\perp,1}$ and $u_{\parallel,1}$) are set at a distance of 10 mm to the bottom plate ($z = 1/4$ $H$). The gray lines shown in Fig. 1(b) mark the respective measuring lines. The ultrasound Doppler method provides instantaneous profiles of the velocity component projected onto each measurement line, $v_x$ ($x$, $t$) measured by $u_{\perp,1-3}$ and $v_y$ ($y$, $t$) measured by $u_{\parallel,1-3}$, respectively. Further detailed information about the experimental setup can be found in [22].

## III. RESULTS AND DISCUSSION

### A. Description of the flow structures, flow regimes

Selected areas of the parameter space $Ra$-$Q$ have already been scanned by a number of previous studies primarily with respect to the formation of different flow regimes. A recent regime diagram can be found in [31]. In the parameter range on which we focus in this study, the system preferably tends to form five convection rolls aligned parallel to the magnetic field, whereby the diameter of the rolls corresponds to the height of the fluid layer. Thus, the number of rolls is naturally predetermined by the aspect ratio $\Gamma = 5$ of the vessel. However, it is also known that the characteristic length scale of the flow structure and thus the roll size can vary under the influence of both $Ra$ and $Q$, in particular, the length scale increases with Ra [32,42], while an increase of the magnetic field reduces the roll size [17,22]. In accordance with this, we observe in our experiments a transition to a 4-roll regime with decreasing $Q$, while 6- or 7-roll structures are found at very high magnetic fields. All quantitative analyses and evaluations in this study are limited to the 5-roll flow pattern.

On the basis of this five-roll structure, we distinguish between three flow regimes, which can be adjusted by varying $Ra$ and $Q$. Figure 2 presents the flow regime diagram containing the main flow structures identified in the measured data:

- Steady regime of quasi-two-dimensional convection rolls
- Oscillating flow regime of quasi-two-dimensional convection rolls
- Oscillating flow regime of three-dimensional convection rolls

The criteria for this classification and a comprehensive description of the respective regimes are given in the following sections. In the first part, we will take a closer look at the structures of the stationary regime before we will analyze the nature of the oscillations in the following sections.

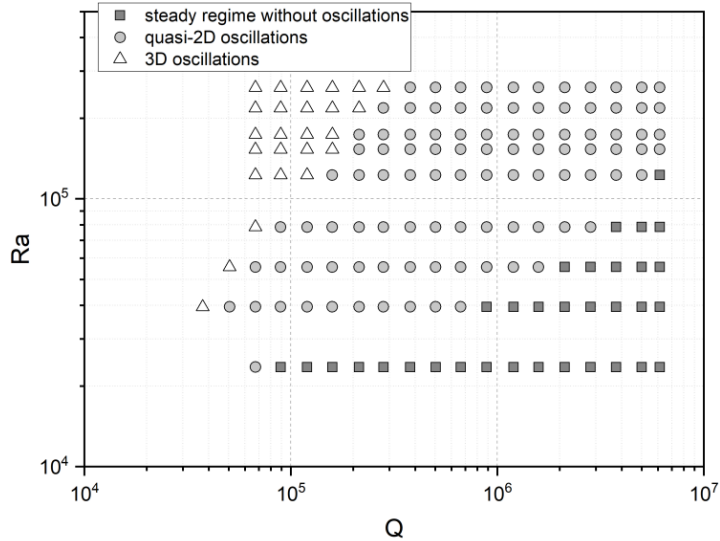

FIG. 2. Flow regime diagram representing the parameter range $Ra$-$Q$ covered within this study

## B. Steady flow

We start our considerations in the range of small $Ra$ numbers and sufficiently large $Q$, in which the magnetic field influence is sufficiently dominant that, although convection occurs, the convection rolls are stationary and are not subject to any temporal fluctuations. The flow is completely re-laminarized by the electromagnetic damping. A corresponding exemplary flow structure is shown in Fig. 3 for a Rayleigh number $Ra = 2.3 \times 10^5$ and a Chandrasekhar number $Q = 5.0 \times 10^6$ ($Q/Ra = 212$).

The spatio-temporal plot in Fig. 3(a) is composed of a time sequence of multiple linear velocity profiles recorded by UDV sensor $u_{\perp,2}$ (see Fig. 1) in the direction perpendicular to the axes of the convection rolls. The ordinate of the diagram represents the $y$-coordinate along the measuring depth, while the abscissa corresponds to the time axis. The colors red and blue represent the $y$-component of velocity in the direction away from or towards the sensor, respectively. From these illustrations the structure of five counter-rotating rolls can be easily identified. Figure 3(b) contains the corresponding time-averaged velocity profile which has an almost perfect sinusoidal shape, so it can be assumed that the cross-sectional areas of the convection rolls should be circular in good approximation.

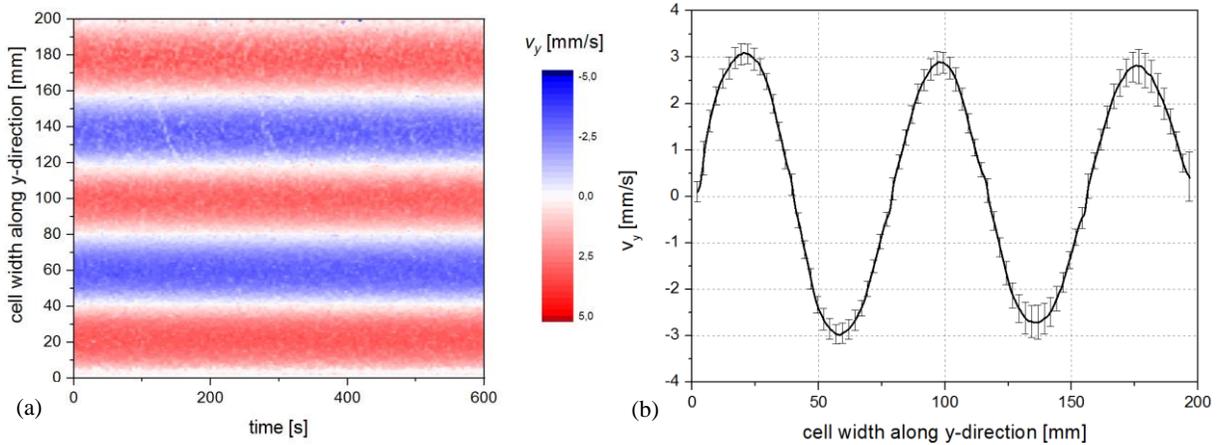

FIG. 3. Flow measurements acquired by sensor $u_{\perp,2}$ at $Ra = 2.3 \times 10^4$, $Q = 5.0 \times 10^6$, $Q/Ra = 212$: (a) Spatio-temporal plot of the flow structure, and (b) time-averaged flow profile $v_y$ on the basis of 320 measured single profiles (error bars are plotted here for every 10th data point and result from the corresponding standard deviations).

A recent study by Vogt et al. [22] revealed a coexistence of the main convection rolls and smaller counter-rotating (secondary) vortex pairs, which are located in the spaces between neighbouring convection rolls at the lid or bottom of the convection cell (see also the schematic representation in Fig. 4(b)). Our UDV sensors are not positioned close enough to the lid or the bottom to measure the smaller side eddies directly. Nevertheless, these substructures leave their footprint also in the time-averaged velocity profiles. Figure 4(a) displays a detail of the velocity profile in Fig. 3(a). Small bulges can be observed at the transition regions between the rolls. This bulging of the lines occur at each zero crossing, which clearly indicates that each gap between the rolls at the lid and the bottom is filled by such recirculating vortices.

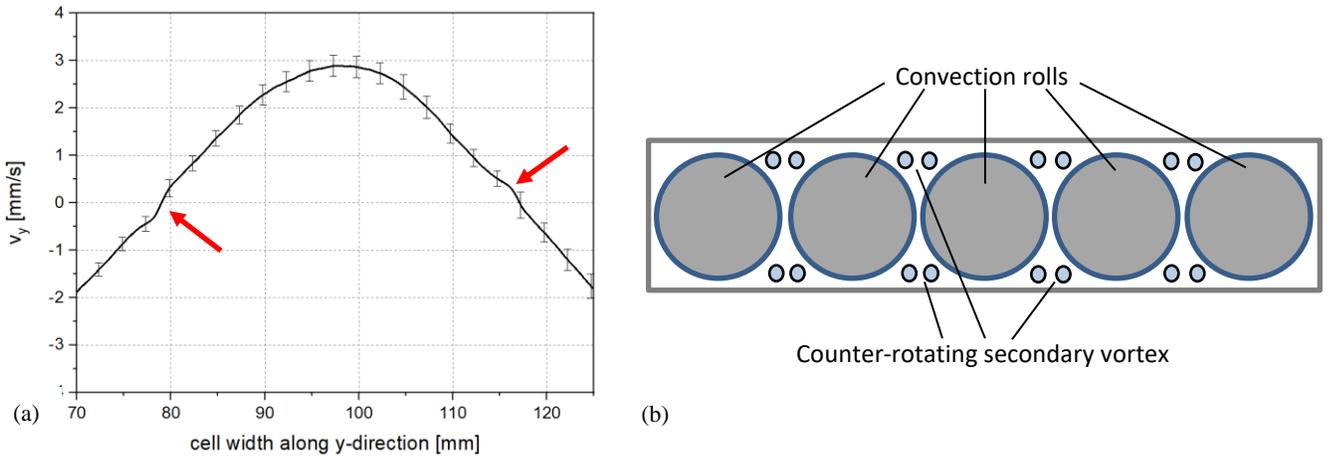

FIG. 4. (a) Detail of the time-averaged velocity profile shown in Fig. 3(a), (b) schematic drawing showing a general illustration of the flow pattern (according to [22])

Figure 5 reveals the influence of varying $Q$ or $Ra$ on the time-averaged profiles of $v_y$. On one hand, a gradual decrease in the magnetic field over a wide range of $Q$ at the small Ra number of $Ra = 3.9 \times 10^4$ is accompanied by an increase in the velocities, but, does not lead to a significant change in the circular roll shape. On the other hand, an increase of $Ra$ obviously causes a distinct deformation of the rolls. It should be noted, however, that this deformation of the rolls does not yet lead directly to time-varying flows. The magnetic field is still sufficiently strong to maintain a steady flow regime and to prevent the onset of oscillatory instabilities. It is also noticeable that the indications for the presence of the secondary vortices apparently disappear at lower values of $Q$. This could have to do with the change in roll size corresponding to modifications of the magnetic field strength. Smaller magnetic fields allow larger rolls, with the consequence that the secondary vortices are pushed back towards the wall or even vanish completely.

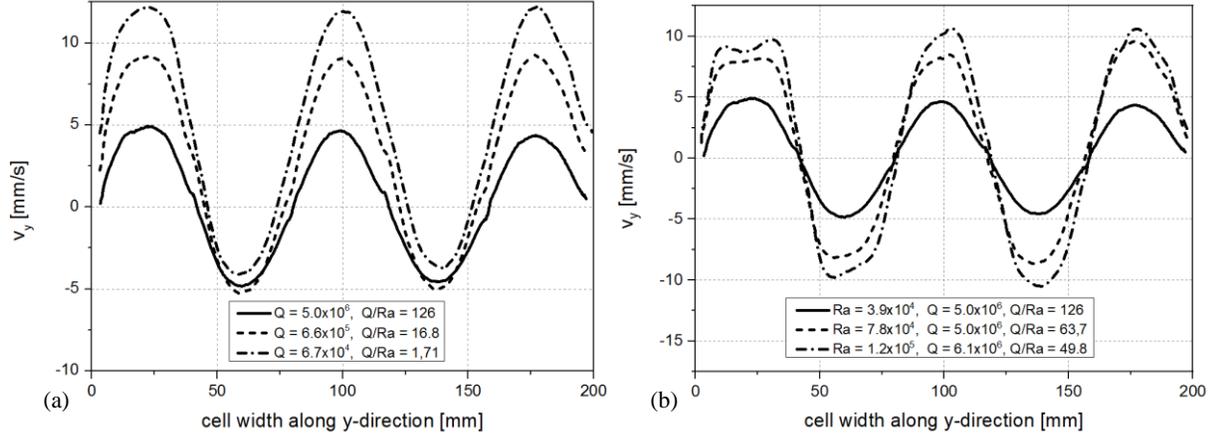

FIG. 5. Time-averaged flow profiles $v_y$ for (a) $Ra = 3.9 \times 10^4$ and varying $Q$, (b) for high $Q$ and varying $Ra$

## C. Quasi-2D oscillations

The transition from a time-stationary to a time-dependent oscillating flow regime occurs when, for a fixed Rayleigh number, the magnetic field strength is lowered so far that the Chandrasekhar number falls below a critical value $Q_{OS}$. If this is the case, distinct and regular oscillations appear in the measurements of both velocity and temperature. We refer to this emerging time-dependent flow as quasi-2D oscillations. An exact description of the flow structure and thus a confirmation of its quasi-2D character are given in this section.

Figure 6 shows a typical example of the velocity and temperature signals resulting from quasi-2D oscillations. The spatio-temporal flow pattern in Fig. 6(a) demonstrates that the three central convection rolls are particularly affected by the oscillations. The side walls parallel to the magnetic field obviously exert a stabilizing influence here. The FFT of the signal time series reveals one dominating frequency. The same frequency was identified by all UDV sensors in all convection rolls and also appears in the temperature measurements as shown in Figs. 6(c) and 6(d). The signals recorded by the thermocouples T1-3, which are situated along the magnetic field lines above the center of the middle roll (see also Fig. 1), oscillate with the same phase. In contrast, the temperature signal at measuring point T4, which is located above the adjacent convection roll, shows a phase shift of 180°. This aspect is also in agreement with the velocity measurements, as a closer look at the flow pattern in Fig. 6(a) shows. The oscillations arise from a periodic and mutual expansion and reduction of the size of neighboring rolls. Consequently, the width of adjacent rolls oscillates in opposite phase. Obviously, we are looking at oscillating roll structures in which the cross-section of the rolls changes periodically. Since the width of the convection cell is unchangeable, the growth or shrinkage of one convection roll is always at the expense of the adjacent rolls.

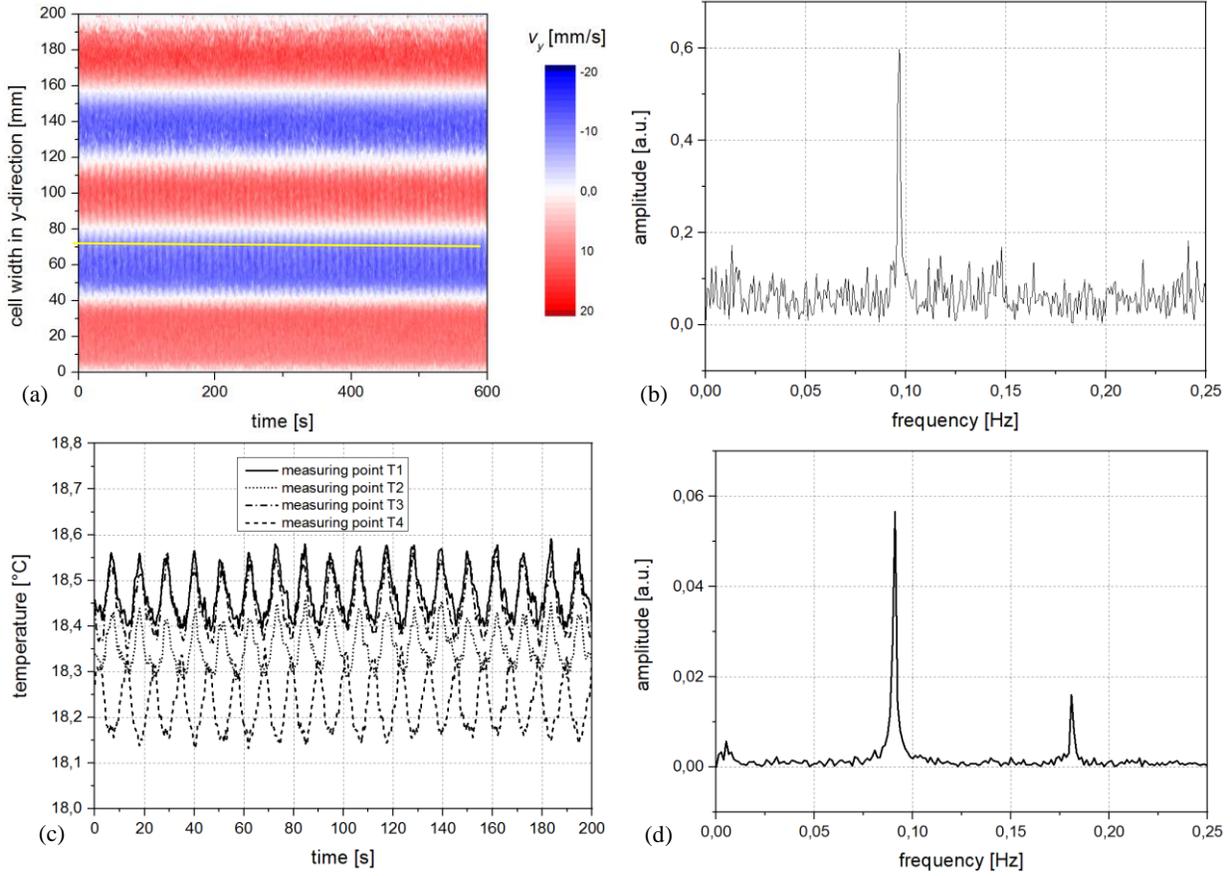

FIG. 6. Two-dimensional oscillations occurring at $Ra = 7.8 \times 10^4$, $Q = 3.8 \times 10^5$, $Q/Ra = 4.82$:

(a) Spatio-temporal plot of the flow structure perpendicular to the field acquired by sensor $u_{\perp,2}$,
(b) Power spectrum of velocity time series obtained by FFT,
(c) Temperature time series,
(d) Power spectrum of temperature time series obtained by FFT

The yellow line marks the position for determining the velocity time series presented in Fig. 8.

It can be assumed that the angular momentum of a convection roll is maintained during the oscillations. This in turn means that its rotation rate would have to vary according to the change in the cross-section. This effect is difficult to measure with the sensors perpendicular to the magnetic field, since the radial measuring position in the roll changes with the roll size. However, it is well-known that the primary motion of a finite rotating column is strongly coupled with the Ekman-driven secondary flow and therefore it is to be expected that such oscillations of the rotation rate are reflected in corresponding fluctuations of the secondary flow [43-46]. In fact, oscillations of the same frequency can be detected in the velocity measurements along the magnetic field direction. Figure 7(a) shows such a typical example of an oscillating pattern of the secondary flow in $x$-direction recorded by UDV sensor $u_{//,2}$.

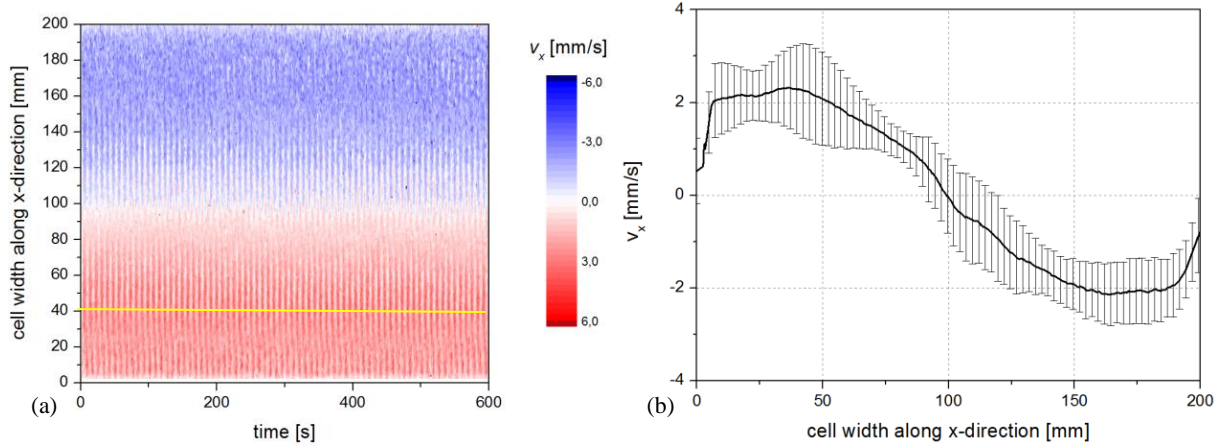

FIG. 7. Flow measurements parallel to the field acquired by sensor $u_{//,2}$ at $Ra = 1.7\times10^5$, $Q = 8.9\times10^5$, $Q/Ra = 5.11$: (a) Spatio-temporal plot of the flow structure, and (b) time-averaged flow profile $v_x$.
The yellow line marks the position for determining the velocity time series presented in Fig. 8. (error bars are plotted here for every 10th data point and result from the corresponding standard deviations).

For the purpose of a clear presentation of the phase relationship of the measured velocity values of the various sensors, sinusoidal fits of the signals are prepared. The resulting graphs are shown in Fig. 8. Remarkable is the almost perfect synchronicity of the oscillations of either the $x$- or the $y$-component. The alleged phase jump with respect to the signal of sensor $u_{\perp,1}$ is explained in a simple way by the different signs of the velocity in the upper and lower part of the convection roll. There is a real phase offset of about 180° between the two velocity components. This is due to the fact that in the case of an oscillating swirling flow, a permanent exchange of energy takes place between the primary flow (rotation) and the secondary flow (Ekman pumping) [44]. The secondary flow transports angular momentum and forms a closed loop of two toroidal vortices where the fluid is moving toward the midplane of the fluid vessel inside the convection rolls ($u_{//,1}$ and $u_{//,2}$). The return flow occurs within the space between neighboring rolls ($u_{//,3}$). Since this gap is smaller than the roll cross-section, we also measure a higher velocity at the UDV sensor $u_{//,3}$ than at the positions inside the rolls.

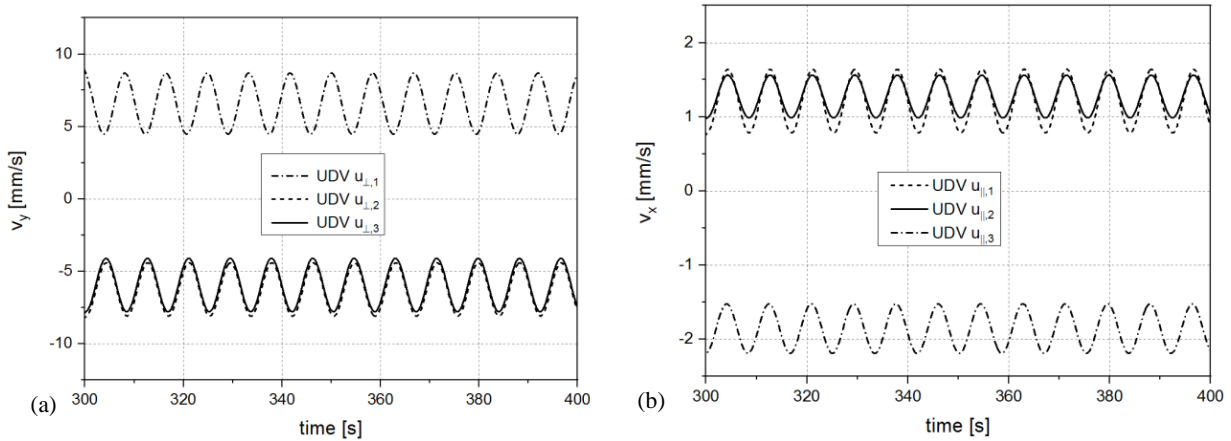

FIG. 8. Sinusoidal fits of time series of velocity recorded at $Ra = 1.7\times10^5$, $Q = 8.9\times10^5$, $Q/Ra = 5.11$: (a) velocity $v_y$ perpendicular to the magnetic field (measuring depth $y = 72$ mm), (b) velocity $v_x$ parallel to the magnetic field (measuring depth $x = 40$ mm). The measuring positions are highlighted in Fig. 6(a) and Fig. 7(a), respectively.

The quasi-2D oscillations observed here are obviously associated with a periodic change of the roll's cross-sectional area. The data suggest that the size of a convection roll changes (either grows or reduces) in all observable directions simultaneously, this phenomenon could therefore be described as a kind of "breathing" flow structure. This also becomes visible when looking at time-shifted snapshots of velocity profiles in y-direction as depicted in Fig. 9 for the UDV sensors $u_{\perp,2}$ and $u_{\perp,1}$. In the example shown here, differences of up to approx. 1 to 2 mm/s can be detected at the respective measuring positions. The UDV measurements achieve a velocity resolution of 0.1…0.2 mm/s, this means that the measurable effect is well above the measurement uncertainty. At an amplitude of approx. 10 mm/s, the differences arising from the "breathing" are roughly estimated to be up to about 10 to 20% of the azimuthal velocity in the convection rolls.

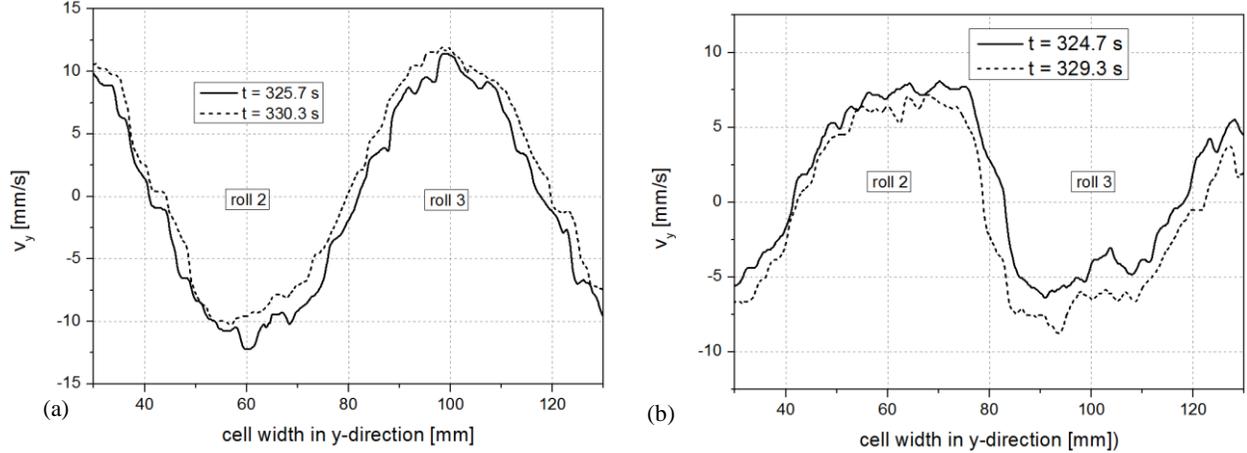

FIG. 9. Sections of instantaneous velocity profiles perpendicular to the magnetic field showing the varying roll size at different points in time: (a) UDV $u_{\perp,2}$, (b) UDV $u_{\perp,1}$. The labels roll2 and roll3 designate the second and third roll seen from the sensor position.

The measurements carried out within this study provide an insight into the topology of the various flow regimes. This detailed view also allows a more precise definition and certain corrections of previous assumptions about the exact flow patterns, the mechanisms for the onset of oscillations and the formation of 3D flow structures. For RBC in a horizontal fluid layer it is widely accepted that, when a critical value $Ra_c$ is exceeded, convection in the form of stationary 2D convection rolls is initiated. An increase of $Ra$ leads to the border of the "Busse balloon" being exceeded, beyond which oscillatory instabilities set in, establishing themselves as traveling waves or modulated traveling waves [1,9,19]. For low $Pr$ fluids such as liquid metals, for which the "Busse balloon" is quite small, the application of a magnetic field shifts the stability limit towards higher $Ra$. Our current findings now reveal that a magnetic field also prevents the development of traveling waves as a first oscillatory instability. If $Q$ is gradually lowered at a fixed $Ra$ number, flow oscillations occur which are really two-dimensional in a strict sense, i.e. no indications of any gradients in the velocity field along the magnetic field lines are found. Our measurements but also earlier investigations show that the convection rolls often do not have an ideal circular cross-sectional area [22,47]. This might suggest that the oscillations could arise from a synchronous rotation of the deformed rolls. However, our analysis does not find sufficient evidence of this assumption. First of all, it is striking that the occurrence of oscillations is not closely related to a certain level of roll deformation. Significant roll deformations are clearly visible in the range of steady flows, while oscillations of almost circular rolls are observed, especially in the range of small $Ra$ numbers.

In the case of the rotation of a deformed roll, one would expect to detect related deviations between the velocity signals of sensors $u_{\perp,1}$ and $u_{\perp,2}$, which are arranged one above the other at the central position of the convection cell perpendicular to the magnetic field (see also Fig. 1). However, this is contradicted by the results found here, for example in Fig. 8(a) showing the sine fits of the time series. It becomes obvious that both signals are exactly in phase. In addition, Fig. 9 reveals a periodic change in the roll size, whereby this also occurs simultaneously in the upper and lower part of the roll. This means that expansion or shrinkage of the rolls occurs simultaneously at different positions. We have already referred to this in the text above with the term "breathing". This is also confirmed by the measurements of sensors $u_{\parallel,1}$ and $u_{\parallel,2}$ parallel to the magnetic field. The oscillations in both signals are almost identical (see Fig. 8(b). Thus, the pronounced fluctuations that can be seen here do not appear to be caused by rotating non-circular rolls, as differences between the sensors should then also occur here. However, this does not exclude the possibility that the movement of some kind of elliptical or even more complicated deformations may occur in this configuration. Rotations of elliptical deformations have been observed, but then the topology of the flow is no longer 2D but has already entered a 3D state, as we will see in the following section.

### D. Transition to 3D oscillations

As soon as measurable deviations in the flow structure along the magnetic field can be determined with decreasing magnetic field by means of UDV or temperature measurements, we proceed on the assumption that three-dimensional disturbances develop and thus the quasi-two-dimensional state no longer exists. A comparison of various measurements has shown that the spatio-temporal plots hardly disclose whether the oscillations can be classified as 2D or 3D structures and when exactly a transition between these regimes occurs. Therefore, at this point we first focus on the detection of potential phase shifts in both the velocity and temperature signals. Figure 10(a) presents sinusoidal fits of time series of velocity obtained at $Ra = 1.7 \times 10^5$, $Q = 1.2 \times 10^5$ ($Q/Ra = 0.69$). A closer look reveals a small but clearly measurable phase shift between the signals of sensors $u_{\perp,2}$ and $u_{\perp,3}$. Furthermore, the mean velocities of both sensors show a significant difference. This already points to a perturbation of the two-dimensional structure. Figure 10(b) shows a comparison of the time-averaged velocity profiles measured by $u_{\perp,2}$ and $u_{\perp,3}$. The diagram reveals a slightly higher velocity in the core of the convection cell at $u_{\perp,2}$ and a displacement of the roll positions, the latter being an unambiguous indication that the convection rolls are no longer aligned exactly parallel to the magnetic field. However, this also means that the flow is no longer quasi-2D, but instead a 3D structure with detectable velocity gradients in the magnetic field direction has been established. The magnetic field is now apparently no longer strong enough to fully compensate for three-dimensional disturbances by momentum transport along the field lines.

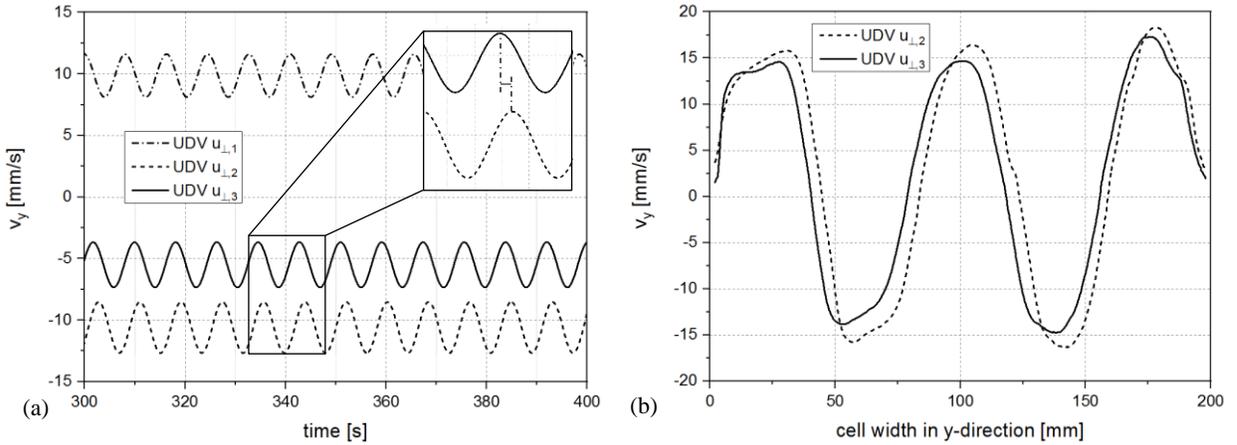

FIG. 10. Characteristics of the velocity perpendicular to the magnetic field as first signs of a transition from a quasi-2D to a 3D flow recorded at $Ra = 1.7 \times 10^5$, $Q = 1.2 \times 10^5$, $Q/Ra = 0.69$: (a) Sinusoidal fits of time series of velocity $v_y$ perpendicular to the magnetic field (measuring depth $y = 72$ mm) (the measuring position is highlighted in Fig. 6(a)), (b) comparison of the the time-averaged velocity profiles, $v_y$, for $u_{\perp,2}$ and $u_{\perp,3}$.

Next, we look at the flow component $v_x$ parallel to the magnetic field. Figure 11(a) presents the secondary flow measured by $u_{//,2}$ which looks almost identical compared to the quasi-2D case shown in Fig. 7(a). Moreover, the comparison between the sensors $u_{//,1-3}$ shown in Fig. 11(c) does not reveal a phase shift. The mean velocities of $u_{//,1}$ and $u_{//,2}$ both monitoring the secondary flow inside the convection roll, differ only slightly from each other. However, a first evidence for a transition to a 3D flow is provided by the measurements made by $u_{//,3}$ in the gap between the convection rolls (see Fig. 11(b)). Considering the $x$- direction parallel to the magnetic field 3D perturbations obviously develop in the center of the convection cell. This is the place where a diverging flow occurs as this is roughly the separation line between the two rolls of the secondary flow. It is obvious that the disturbances affect and deform the smaller recirculating vortex pairs, which fill the space between the rolls [22]. These deformations are transported by the secondary flow from the center towards the side walls. This is also clearly reflected in the FFT spectrum, in which new low-frequency peaks appear (see Fig. 11(d)). Our analysis has proven that the frequency of these very long-wave fluctuations $f_{lw}$ corresponds almost exactly to the turnover frequency of the secondary flow, $f_{lw}/f_{to\_ekman}$ ≈ 1. Thus it is clear that we see here the advection of three-dimensional disturbances by the secondary flow along the direction of the magnetic field. A very similar behavior is known from a flow in a cylinder driven by a rotating magnetic field. In that case, evolving instabilities occur in form of so-called Taylor-Görtler vortices along the inner side walls at the mid-height of the container [22,45,48]. In this case, the transition to a three-dimensional flow results from a splitting, tilting and advection of the Taylor-Görtler vortex pair [49].

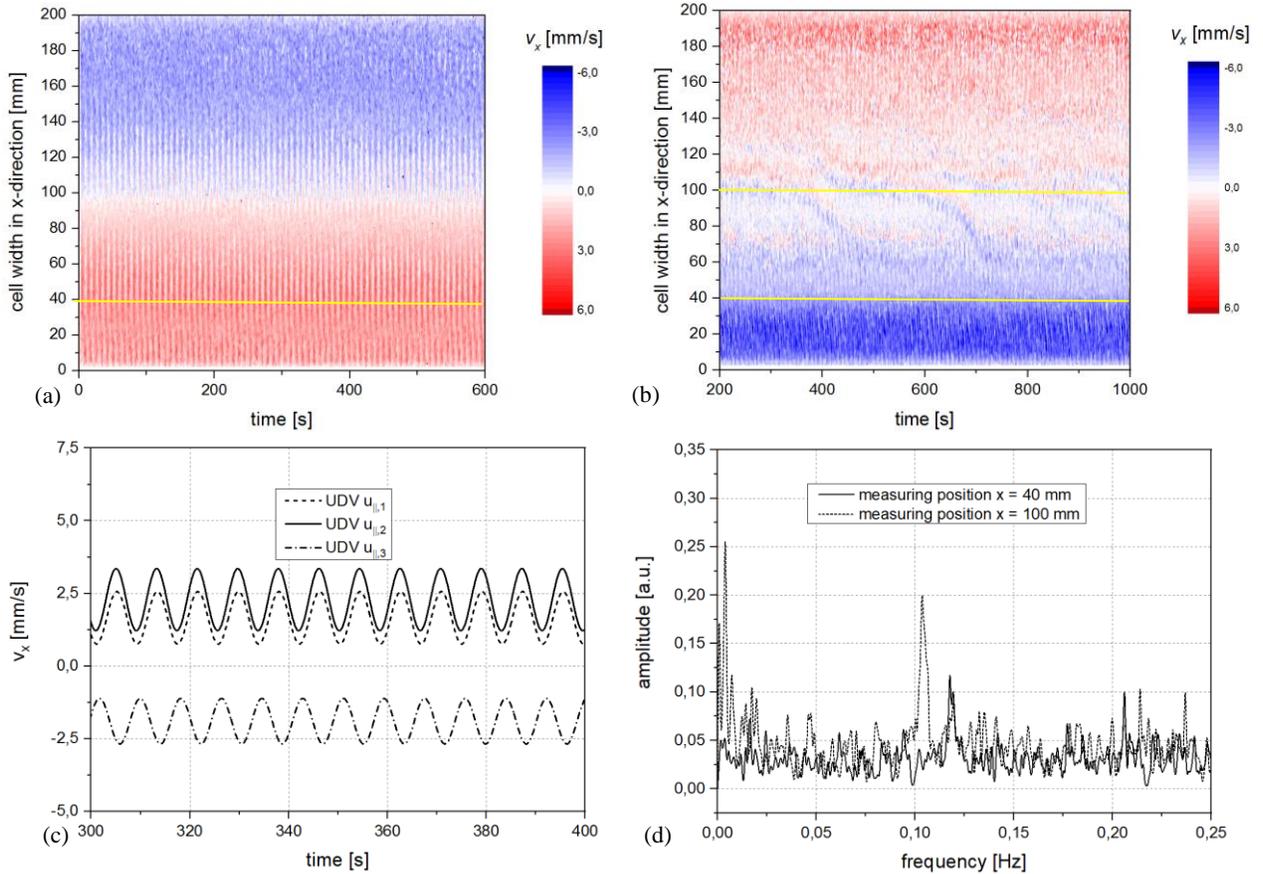

FIG. 11. Characteristic properties of the velocity parallel to the magnetic field at $Ra = 1.7 \times 10^5$, $Q = 1.2 \times 10^5$, $Q/Ra = 0.69$: (a) Spatio-temporal plot of the flow structure measured by sensor $u_{//,2}$, (b) spatio-temporal plot of the flow structure measured by sensor $u_{//,3}$, (c) Sinusoidal fits of time series of velocity $v_x$ parallel to the magnetic field (measuring depth $x = 40$ mm), (d). Power spectrum of velocity time series obtained by FFT for sensor $u_{//,3}$ at measuring positions $x = 40$ mm and $x = 100$ mm. Both measuring positions are highlighted in (a) and (b).

## E. Effects on the Re number and oscillation frequency

Regarding the presentation of results in the previous sections, special emphasis was put on the description of the topology of the individual flow regimes. This section is devoted to the behavior of the global parameter Reynolds number $Re$ in the different flow regimes shown in Fig. 2. In this study we define the Reynolds number as $Re = v_{y\_max} H / \nu$, where the characteristic velocity $v_{y\_max}$ is the mean value of the time-averaged maximum velocity in all five convection rolls measured by sensor $u_{\perp,2}$. Figure 12(a) displays these velocity values for various $Ra$ numbers as a function of $Q$. The same data are presented in Fig. 12(b) in the form of the compensated $Re$ number $Re \cdot Ra^{-1/2} = f(Q/Ra)$. In this graph all curves converge fairly well. It is obvious that there is no monotonic dependence on $Q$. In fact, three different domains can be identified. This becomes particularly evident in detail in Fig. 12(c). In zone I, which corresponds to the application of weak magnetic fields, an increasing magnetic field strength leads to an increase in flow intensity. After passing a maximum value, a reduction in the rotation speed of the convection rolls with increasing $Q$ can be observed in accordance with common expectations. This range is then further subdivided into zones II and III, which differ in the deviating decrease of $Re$ with $Q$. At large values of $Q/Ra$ the steepening slope of the curves apparently approaches a power law of $(Q/Ra)^{-1/4}$. Linear fits for $Ra = 2.3 \times 10^4$ and $Ra = 3.9 \times 10^4$ yield exponents of -0.236 and -0.241, respectively. The determination of the exponents for larger $Ra$ is not feasible, since it is not possible to achieve sufficiently high values $Q/Ra$.

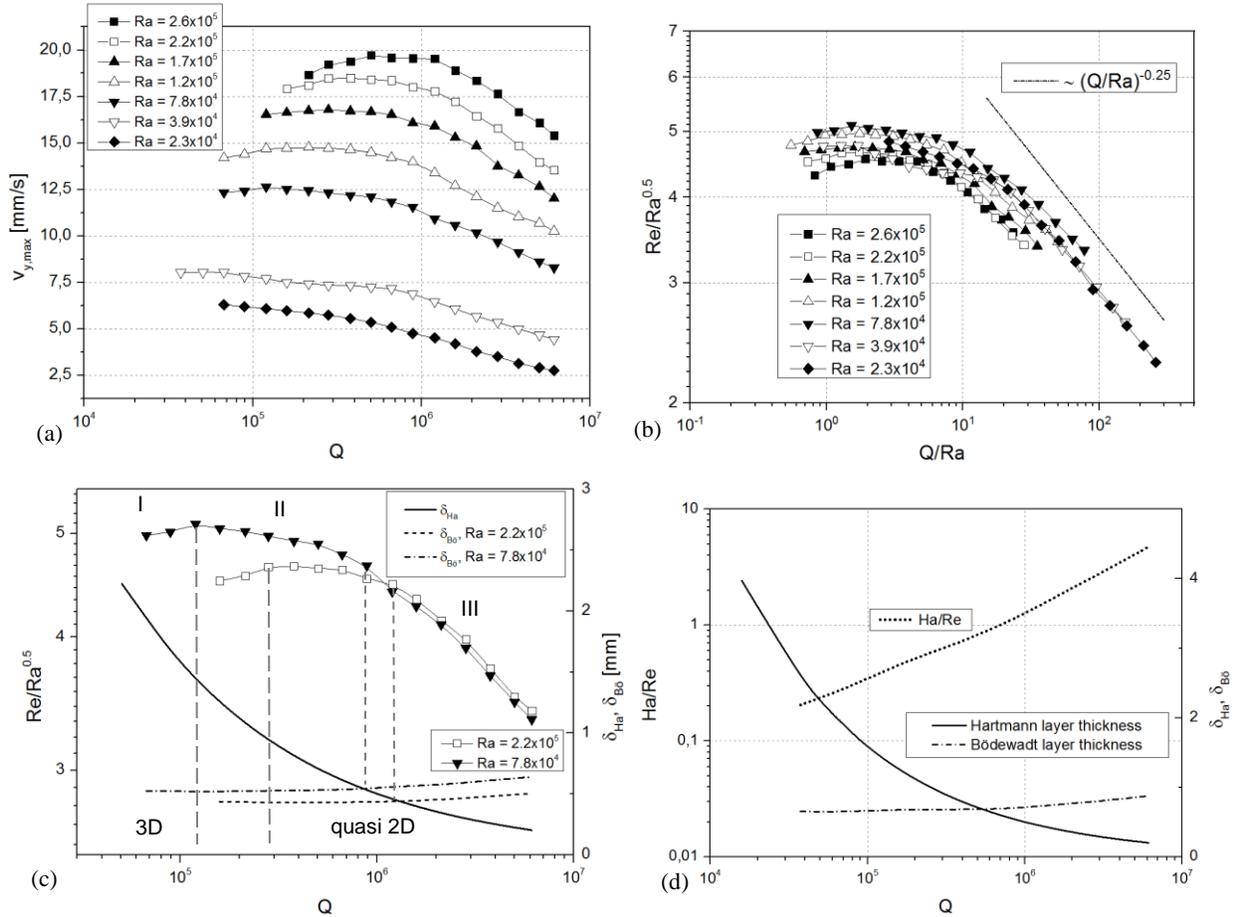

FIG. 12. (a) Time-averaged maximum velocity of the convection rolls as a function of $Q$, (b) corresponding values of the compensated $Re$ number, $Re/Ra^{0.5}$, plotted vs. $Q/Ra$, (c) comparison of $Q$-dependencies of the compensated $Re$ number and the extension of both the Hartmann and Bödewadt layers (d) comparison of $Q$-dependencies of $Ha/Re$ and the boundary layer thicknesses for $Ra = 7.8 \times 10^4$.

It is interesting to note that the location of the maximum $Re$ number corresponds roughly to the parameters where the transition between quasi-2D (zones II and III) and 3D flow is observed. In [23] it was shown that the application of the magnetic field to an unstable roll regime leads to an increase of the velocity component perpendicular to the magnetic lines. From this one can infer that the energy transport in favour of the main convection rolls will continue as long as the flow has not completely reached the quasi-2D state. Once the two-dimensional regime is established, any further increase in field strength results in a decay of the flow. In this context it remains to be clarified why two different slopes are observed for the zones II and III. The situation considered here is characterized by the fact that two boundary layers exist at the wall perpendicular to the magnetic field, also known as Hartmann walls, the Hartmann layer $\delta_{Ha} = L/(2Ha) = L/(2Q^{0.5})$ and the Bödewadt layer $\delta_{B\ddot{o}} = \sqrt{\nu/\Omega}$ [22,30]. If $Q$ is sufficiently small, the Hartmann layer is considerably thicker than the Bödewadt layer, so that the viscous boundary layer is determined by a balance between viscous and inertial forces. As $Q$ increases, $\delta_{Ha}$ will at some point fall below $\delta_{B\ddot{o}}$, the Lorentz force becomes dominant and the thinner Hartmann layer is henceforth to be decisive for the viscous damping. Figure 12(c) shows for two selected $Ra$ numbers that the crossing point of the graphs for $\delta_{Ha}$ and $\delta_{B\ddot{o}}$ is detected at exactly those values of $Q/Ra$ where one could define the transition from zone II to zone III. Thus the two different slopes can obviously be assigned to different damping mechanisms. This agreement is also observed for all other $Ra$ considered in this study. It is also found that at the same $Q$ the ratio $Ha/Re$ reaches approximately a value of one (see Fig. 12(d)). It is well-known that the damping of quasi-two-dimensional MHD flows in a finite vessel exposed to a strong magnetic field (high $Q$ range) is quite low and governed by the Hartmann braking only [18]. Knaepen and Moreau [29] suggest that $Ha/Re$ is the relevant parameter to characterize the Hartmann damping, whereby $Ha/Re \approx 1$ means that both the Hartmann damping time, $\tau_H$, and the typical turnover time of the flow structure, $\tau_{tu}$, are of the same order of magnitude.

The increase of the $Re$ number with increasing $Q$ for the case of the 3D flow is noteworthy and should be discussed here briefly. Various studies considering thermally driven convection under the impact of various stabilizing forces such as rotation, geometrical confinements or static magnetic fields report an enhancement of scalar transport by applying moderate forces [50]. Tagawa and Ozoe [51] considered the vertical convection in a cubic container under the influence of a horizontal magnetic field applied parallel to the tempered walls. Their numerical results predict an improvement of convective heat transfer for a wide range of Rayleigh numbers by applying a weak magnetic field, while Nusselt number decreases with increasing field strength for strong magnetic fields. The numerical predictions were verified in experiments where Nu was determined as a function of the field strength. Similar results were found in experimental and numerical studies done by Burr et al. [52], Tagawa et al. [53] and Authie et al. [54] considering buoyant flows in long vertical enclosures in a horizontal magnetic field perpendicularly aligned to the horizontal temperature gradient. A detailed study dealing with the effect of a horizontal magnetic field on Rayleigh-Bénard convection was presented by Burr and Müller [17]. A range of moderate Ha numbers was likewise identified, in which a distinct increase in heat transport could be observed in comparison to the case without magnetic field. Their work does not include direct measurements of the flow structure, but some qualitative conclusions were drawn from the temperature data. Local isotropy coefficients of the local temperature gradient indicate that the increase of the heat transfer is associated with an increasingly non-isotropic state of the convection rolls. Any 3D flow under the influence of a magnetic field is subjected to Joule damping, which is much stronger than the Hartmann braking that controls the emerging 2D structures. Nevertheless, an increase of $Re$ with $Q$ is observed especially in this range (zone I in Fig. 12(c)), which can only be explained by the fact that the damping effect of the Joule dissipation on the convective motion is counterbalanced by an ordering effect of the magnetic field which may improve the heat transfer. It is assumed that an increase in the degree of coherence of the global flow in the convection cell, which essentially determines the transport properties, leads to this result [50,55]. Our results show that with the occurrence of a quasi-2D flow, the optimal state of a coherent flow is obviously reached. With each further increase of the magnetic field Joule damping is substituted by Hartmann braking, but no further improvement of the flow coherence can be achieved. Consequently, the $Re$ curve passes its maximum when the quasi 2D state is reached and declines in the further course. This optimal state of coherent flow correlates with a maximum heat transport [23].

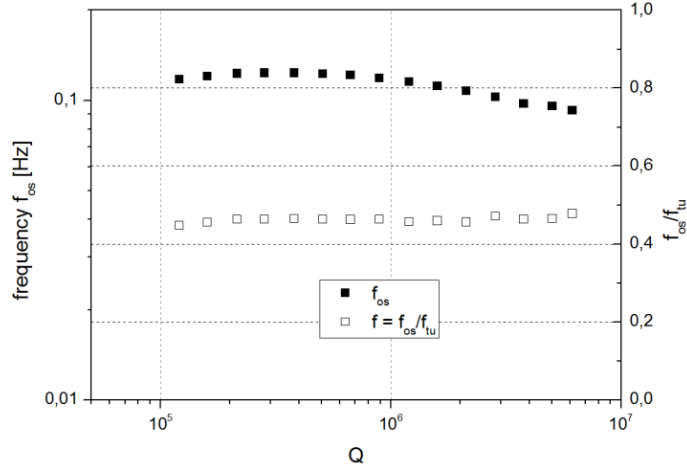

FIG. 13. Dominant frequency of the oscillations $f_{os}$ and the ratio $f_{os}/f_{tu}$ vs. $Q$ for a selected $Ra$ number $Ra = 1.7 \times 10^5$

It was already mentioned above that in all oscillating time signals of velocity and temperature the same dominant frequency $f_{os}$ is found, which is plotted in Fig. 13 as a function of $Q$ for $Ra = 1.7 \times 10^5$. This dependence is very similar to that of the mean roll velocity and the $Re$ number, respectively. Figure 13 also reveals that $f_{os}$ is in a close relationship to the turnover frequency $f_{tu}$ which was calculated from the measured velocities $v_y$ and the path of a fluid element along a circular trajectory with the distance of the sensor position from the centre plane of the convection cell as radius $r = 10$ mm. The ratio $f_{os}/f_{tu}$ shows constant values of about 0.45±0.03. This also applies to all other $Ra$ considered within this study. The fact that the oscillation frequency and turn-over frequency do not coincide is a further indication that the observed oscillations are not due to any imbalances or asymmetries of the roll shape in connection with the rotation of the convection rolls. Nevertheless, the question arises regarding the reason for the solid coupling between $f_{os}$ and $f_{tu}$. In section C it was already noted that the oscillations of the convection rolls are reminiscent of a "breathing" structure. This manifests itself in a periodic increase and decrease of the intensity and extension of the rotary motion of the rolls and is associated with corresponding angular momentum oscillations. These fluctuations of angular momentum need to be compensated by the secondary flow parallel to the magnetic field. Accordingly, we observe closely correlated velocity fluctuations in the convection rolls as well as in the secondary flow (see Figs. 6 and 7). Apart from the difference in terms of the flow driving mechanism, there are remarkable similarities between the thermal convection considered here and the swirling flow in a cylindrical column driven by a rotating magnetic field (RMF). Previous studies on the latter topic have shown that disturbances of various kinds can trigger the development of pronounced oscillations in a rotating flow structure and that these oscillations can be described as standing inertial waves [44,46]. A striking feature of inertial waves is the limitation that their frequency $f_{iw}$ is always less than twice the rotation frequency of the basic flow with angular velocity $\Omega$:

$$f_{iw} \leq \Omega/\pi \qquad [1]$$

which is equivalent to the requirement $f_{os}/f_{tu} \leq 2$. An inertial mode is characterized by wave numbers in radial ($\gamma_i$), axial (n) and azimuthal direction (m). According to our velocity measurements we can assume that the inertial wave is axisymmetric (m=0) and appears as ($\gamma_1$,n=2) mode. For m = 0, $\gamma_1 = 3.832$ is the root of a first-order Bessel function and for an estimation of the frequency $f_{\gamma_i,n}$, we can apply the linear theory for inviscid inertial waves in a straight circular cylinder [56]:

$$f_{\gamma_i}n = \frac{\Omega}{\pi} \left[ 1 + \left( \frac{\gamma_i}{n\pi} \right)^2 \left( \frac{L}{R} \right)^2 \right]^{-1/2} \qquad [2]$$

With $L/R = 2\Gamma = 10$ we obtain $f_{\gamma_1,2} / f_{tu} \approx 0.33$. Although this value is lower than that measured in this study, this approach also establishes a solid and plausible link between $f_{os}$ and $f_{tu}$. This moderate difference can be explained by the fact that the Rayleigh-Benard setup deviates significantly from the assumptions of the linear inertial wave theory, in which the flow resembles a solid body rotation in an inviscid fluid under no-slip boundary conditions. Furthermore, the effect of a DC magnetic field on the dynamics of inertial waves has not been investigated yet.

### F.  Transitions in the parameter space

This study illuminates a limited section of the $Ra$-$Q$ parameter space for a certain $Pr$ number of 0.03 and a fixed aspect ratio of five. The respective regime diagram shown in Fig. 2 (Sec. III.A.) reveals the existence of three different flow regimes in this area: steady regime of quasi-2D convection rolls, and two time-dependent oscillatory regimes. The area of the steady flow corresponds to the "Busse balloon" which is remarkably extended here due to the strong horizontal magnetic field. The two borderlines between these three areas can actually already be deduced from Fig. 2, but at this point we would like to take a closer look at the transitions.

The tendency for the magnetic field to significantly delay the onset of oscillatory instabilities is clearly confirmed by our measurements, as it becomes apparent in Fig. 14(a). Following Busse and Clever [1] we plotted here the difference $Ra_{os}$-$Ra_c$ vs. the stabilizing Chandrasekhar number $Q$. $Ra_{os}$ determines the threshold, where the first oscillatory instability occurs, and $Ra_c$ is the critical Ra number for onset of convection. With the given experimental setup it is not possible to generate the extremely small temperature gradients necessary to determine $Ra_c$ directly from the experiments. We therefore used numerical simulations performed by Yanagisawa et al. [57] for exactly the same experimental configuration considered here, who calculated a value of $Ra_c = 1780$ for the non-magnetic case. Furthermore, we used the relation $Ra_c = 4\pi^2 Q^{1/2}/(L/2)$ given by Burr and Müller [17] for the limit case of very high $Q$ numbers for non-conductive walls. Although we map a significantly higher $Q$-range as those covered by Busse and Clever [1], the observed dependence on Q appears to be very similar. Fauve et al. [16] found in their mercury experiments a power law dependence $Ra_{os}$-$Ra_c \propto Q^{\alpha}$ with $\alpha = 1.2$. Thus, we have added a respective function graph $Ra \propto Q^{1.2}$ to the diagram in Fig. 14(a). It becomes obvious that our data approximate this power law for large values of $Q$. Figure 14(a) also contains a comparison to corresponding data published by Burr and Müller [17]. In their paper, Burr and Müller [17] defined $Q$ with the height $H$ of the convection cell, while we used the dimension $L/2$ of the cell in the direction of the magnetic field. To make both data sets comparable, we adjusted the $Q$ values of Burr & Müller (2002) accordingly. Nevertheless, a considerable difference must be noted, Burr and Müller [17] detected oscillations in their temperature signals at much lower Ra numbers (see the full triangle symbols). The main difference between both experiments consists in the dimensions of the convection cells used, in particular the aspect ratio which is $\Gamma = 5$ and 20, respectively. It is known from previous studies that the aspect ratio has a significant influence on the onset of convection (see for instance Stork and Müller [58]). The presence of lateral walls has a stabilizing effect, the stability region increases as the aspect ratio decreases. Fauve et al. [16] carried out comparable experiments in two separate boxes with aspect ratios of 4 and 6 ($\Gamma_1/\Gamma_2 = 1.5$), respectively. In their case, the values for $Ra_{os}$-$Ra_c$ associated with the transition from steady state to oscillatory flow differ by a factor of approximately 4. If we want to compare our data with those in [17], a larger difference in the aspect ratios ($\Gamma_1 = 20$ in [17], $\Gamma_2 = 5$ in this study) must be taken into account resulting in the factor $\Gamma_1/\Gamma_2 = 4$. Assuming a simple proportionality regarding the effect of the aspect ratio on the critical $Ra$ number, it can be roughly estimated that our data should differ by a factor of about 10 from those of Burr and Müller [17]. An appropriate rescaling does indeed lead to a fairly good agreement (see the open triangle symbols).

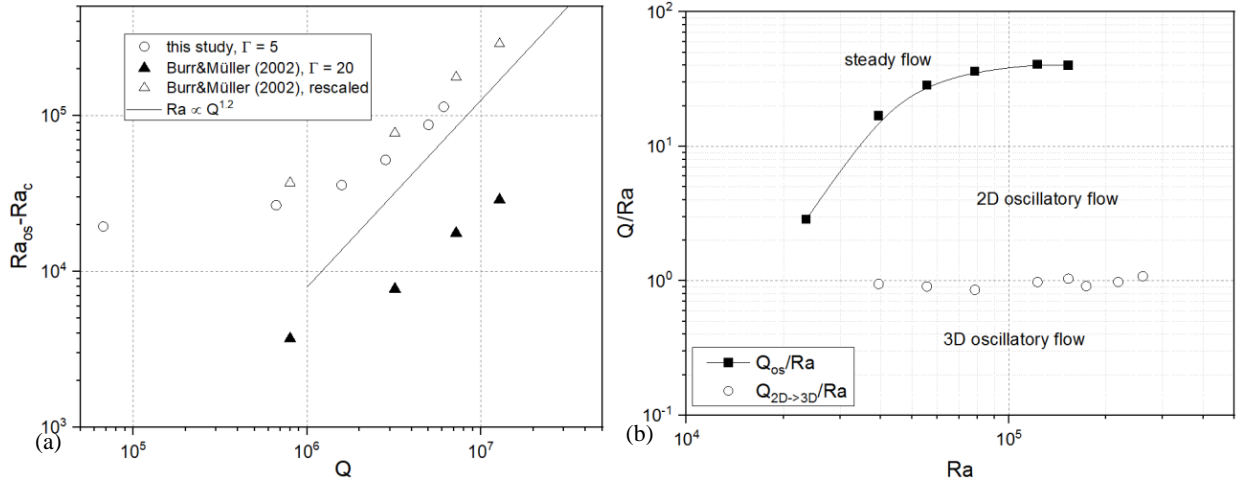

FIG. 14: Transitions in the parameter space: (a) critical parameter $Ra_{os}$-$Ra_c$ for the onset of oscillatory convection, (b) critical values $Q/Ra$ for both transitions from steady to oscillatory flow and from quasi-2D to 3D oscillations.

The electromagnetic forces clearly dominate the momentum equation and inhibit the occurrence of time-dependent flows for very large values of $Q$, i.e. $Q/Ra \gg 1$. Figure 14(b) presents the critical values $Q/Ra$ for both transitions from steady to oscillatory flow ($Q_{os}$) and from quasi-2D to 3D oscillations ($Q_{2D \to 3D}$). If drawing the value $Q_{os}/Ra$ as a function of $Ra$, it appears to converge towards a value of about 40 for large $Ra$. It is interesting to see that the transition from 2D to 3D flow structures is observed especially at values of $Q/Ra \approx 1$. After we have seen, however, what great influence the aspect ratio has on the onset of convection, it would certainly be somewhat premature to regard this result as generally valid. For this purpose, further investigations considering different aspect ratios would be highly desirable.

## I.V. Summary and conclusions

This experimental study concerns the effect of a strong horizontal magnetic field on the Rayleigh-Bénard convection in a horizontal liquid metal layer. The nature of magnetic dissipation is responsible for the formation and stabilization of two-dimensional convection rolls with their axes parallel to the magnetic field direction. The detailed evaluation of velocity and temperature data provides a descriptive picture of the development of the flow, beginning with a stationary roll regime at high $Q$, via the onset of two-dimensional oscillations with decreasing magnetic field strength, and ending with the emergence of three-dimensional instabilities. Furthermore, the focus is limited to a flow regime of five counter-rotating convection rolls, which naturally dominates in a wide parameter range of $Ra$ and $Q$ numbers in such a convection cell with a rectangular geometry and a given aspect ratio of $\Gamma = 5$.

The quasi-2D flow regimes (both the steady rolls as well as the quasi 2D oscillations) are characterized by the following features: (i) The primary flow occurs in the $y$-$z$-plane and does not show any velocity gradients in the direction of the magnetic field, except in the Hartmann and Bödewadt boundary layers, respectively, (ii) a secondary flow is driven by Ekman pumping composed of two toroidal vortices along each convection roll, (iii) there is no phase shift in the velocity and temperature oscillations along the magnetic field lines, and (iv) additional smaller vortex pairs exist in the interspace between the main convection rolls.

Deformation of the roll's cross-section, in particular, a deviation from a circular cross-section are found at higher $Ra$ numbers. However, this alone is not yet a sufficient reason for the transition to an oscillating flow regime. This conclusion is based on the following findings:

(a) Oscillations also occur at small $Ra$ numbers ($Ra \leq 3.9 \times 10^4$), where we do not yet observe any significant deviations from the circular shape of the rolls,

(b) a sufficiently strong magnetic field is able to suppress the oscillations at moderate $Ra$ number (for instance at $Ra = 7.8 \times 10^4$, see also fig. 2), but without restoring a circular roll shape, and

(c) the deformation of the rolls significantly depends on $Ra$, while the ratio of oscillation frequency and turnover frequency $f_{os}/f_{tu}$ remains almost constant over the whole parameter range.

The concept of the 'Busse balloon' predicts the existence of a stability region just above the onset of convection characterized by a flow regime of stationary two-dimensional convection rolls. With an increase in Ra, the boundary of this range is exceeded at some point, beyond which a time-dependent flow appears. The application of a DC magnetic field is supposed to stabilize two-dimensional flow structures along the magnetic field lines and thus extends the area of the 'Busse balloon' [1,10]. With increasing $Ra$ or decreasing magnetic field (decreasing $Q$), these stationary rolls become unstable with respect to an oscillatory flow, which initially also has a quasi-2D structure, i.e. each periodic change of the flow structures occurs synchronously and with the same amplitude along the magnetic field direction. These oscillations are caused by periodic fluctuations of the roll's cross-section resulting from isotropic changes in the roll dimensions perpendicular to their axes. All velocity signals recorded at different positions along the magnetic field direction are synchronous without any phase shift. There is one dominating frequency for all sensors at all positions. This also applies to the temperature data measured along the magnetic field. We have found no evidence that the two-dimensional oscillations are directly related to a deformation of the convection rolls. Our findings suggest that global elliptical oscillations cannot take place without affecting and deforming the smaller vortex structures between the main convection rolls. However, this causes 3D disturbances in these zones, which immediately come under the influence of the secondary flow.

First indications of a deviation from a 2D structure manifest themselves in a beginning tilting of the roll axes with respect to the direction of the field lines shown by a shift of the signals in terms of amplitude and phase and a position-dependent offset of the roll's position. In consequence, there are measureable velocity differences along the magnetic field. Further 3D perturbations originate in the areas between the rolls covered by the recirculating vortex pairs. Velocity measurements made in this zone along the magnetic field (UDV sensor $u_{\parallel,3}$) are a very sensitive indicator to detect such upcoming instabilities. It discloses irregularities of the roll boundaries and the smaller vortex pairs while these disturbances are advected by the secondary flow.

The most obvious effect of the magnetic field is the suppression of 3D flow patterns and the promotion of two-dimensional structures aligned along the magnetic field lines. In line with this, in the 3D regime (zone I in fig. 12(c)) the $Re$ number increases with growing $Q$ while it decreases both in the steady and oscillatory 2D regime. The Lorentz force sets a selection process into operation force preferring less dissipative structures that tend to adopt a two-dimensional state. The strengthening of these structures can be explained by the fact that they receive more energy from the mean flow than they dissipate [14,15]. The parameter combination, where the maximum Reynolds number is measured, corresponds to the transition between 2D and 3D flows and can be considered as an optimal state of coherence in the flow structure. This highly coherent flow is associated with a remarkable increase in the Nu number [23,55]. As soon as the oscillations are quasi-2D, any further increase in the magnetic field only leads to increased dissipation in the Hartmann layers [14]. At this point the intensity of the flow must decrease with further increasing magnetic field.

The comparison of our work with the results of Burr and Müller [17] also showed, however, that the aspect ratio obviously has a significant influence on the onset of convection. Therefore, further investigations considering variations in the aspect ratio would be desirable. Smaller aspect ratios in connection with increased heights $H$ of the convection cell enable higher $Ra$ numbers to be achieved and thus extend the investigations to the range of MHD turbulence.

## Acknowledgement

The authors would like to thank Yuji Tasaka and Takatoshi Yanagisawa, whose suggestions were one of the reasons to initiate this study and who have contributed in fruitful discussions to raise various interesting questions related to this specific topic. The authors also thank Sylvie Su, Felix Schindler and Sanjay Singh for very stimulating and fruitful discussions. The participation of J.C. Y. in this project is financially supported by CSC (China Scholarship Council).